\documentclass{mem}
\usepackage{txfonts}\usepackage{balance}
\usepackage{graphicx}
\usepackage[a4paper,breaklinks,dvipdfm]{}
\idline{84}{1}
\usepackage{txfonts} 

\begin{document}
\def\teff{$T\rm_{eff }$}
\def\kms{$\mathrm {km s}^{-1}$}

\title{
Ultra-fast outflows (aka UFOs) \\ from AGNs and QSOs
}

   \subtitle{}

\author{
M. \,Cappi\inst{1}
\and
F. \,Tombesi\inst{2,3}
\and
M. \,Giustini\inst{4}
          }

  \offprints{M. Cappi}

\institute{
INAF$/$IASF-Bologna, Via Gobetti 101, I-40129, Bologna, Italy
\and
NASA Goddard Space Flight Center, Greenbelt, MD 20771, USA
\and
Department of Astronomy, University of Maryland, College Park, MD 20742, USA
\and
ESAC/ESA, PO Box 78, E-28691 Villanueva de la Canada, Madrid, Spain
\\
\email{massimo.cappi@inaf.it}
}

\authorrunning{Cappi}

\titlerunning{UFOs in AGNs and QSOs}

\abstract{
During the last decade, strong observational evidence has been accumulated for the existence 
of massive, high velocity winds/outflows (aka Ultra Fast Outflows, UFOs) in nearby AGNs and in 
more distant quasars.
Here we briefly review some of the most recent developments in this field and discuss the relevance of UFOs 
for both understanding the physics of accretion disk winds in AGNs, and for quantifying the 
global amount of AGN feedback on the surrounding medium.
\keywords{X-rays: Galaxies -- Galaxies: nuclei -- Galaxies: quasars: absorption lines} 
}
\maketitle{}

\vspace{-2 cm}
\section{Introduction}

Mass outflows/winds have been found in all types of AGNs and QSOs. 
Despite being radiatively dark/silent, they have revealed themselves via blue-shifted absorption features 
imprinted on the source UV and X-ray continua (see Crenshaw, Kraemer \& George 2003 for a review). Once corrected for sample selection biases, intrinsic UV absorption lines 
blue shifted by 10$^{3-5}$ km s$^{-1}$ are generally found in about 50-60\% of all AGNs (Ganguly \& Brotherton 2008, Hamann et al. 2012). 
Similarly, in X-rays, warm absorbers (WAs) have been observed in about 50\% of bright local AGNs and 
more distant QSOs (Reynolds 1997; McKernan et al. 2007; Porquet al. 2004; Piconcelli et al. 2005). 
In the last decade, thanks to the high throughput provided by contemporary satellites such as XMM-Newton, Suzaku and Chandra, 
gas at even higher ionization state (i.e. FeXXV, FeXXVI), column density (N$_W$$\sim$10$^{22-23}$cm$^{-2}$) and with much 
higher velocities (v$\sim$0.05-0.3c) has been observed in a number of AGNs/QSOs at both low and high redshift (Pounds et al. 2003;
Chartas et al. 2002; Tombesi et al. 2010 and references therein).
Here some recent developments in this field are presented (\S 2) with a particular emphasis on their importance for our understanding 
of accretion disk formation/acceleration (\S 3), and their (global) relevance for AGN feedback (\S 4).

\begin{figure*}[t!]
{\includegraphics[width=13cm,height=4cm]{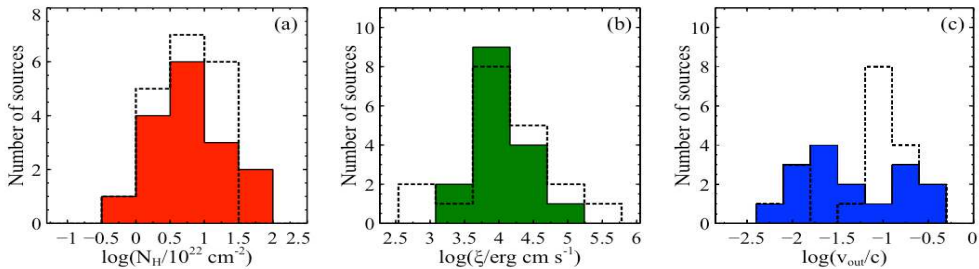}}
\caption{\footnotesize
Distributions of mean absorber parameters: (a) logarithm of the mean column density; (b) logarithm of the mean ionization parameter; 
(c) logarithm of the mean outflow velocity. The red, green and blue colors indicate the Suzaku analysis, while the dotted (black) lines 
are the XMM-Newton analysis (from Gofford et al. 2012).
}
\label{eta}
\end{figure*}

\vspace{-0.3 cm}
\section{Recent Results}

Systematic studies with XMM-Newton have found that as many as $\sim$40\% of local (z$<$0.1) AGNs show evidence for highly-ionized, 
high-velocity Fe absorption lines interpreted as ultra-fast outflows (UFOs) from the central nucleus 
(Tombesi et al. 2010). More recently, a new systematic study has been conducted by Gofford et al. (2013) using also archival data from the 
Suzaku space observatory, allowing for a wider (0.6-50 keV) energy band, a high throughput up to 8-9 keV, several independent instruments 
(XIS0/1/2/3) and a different but comparable sample of (51 type 1-1.9) AGNs. This study was unique in its way to verify, complement and 
possibly expand on the previous XMM-Newton results. Remarkably, the results (see Fig. 1) are entirely consistent with the XMM-Newton ones 
independently providing strong supporting evidence for the existence of very highly-ionised circumnuclear material in a significant fraction of both 
radio-quiet and radio-loud AGN in the local universe.

In fainter, more distant QSOs, X-ray studies have been generally limited by the low statistics. However, after the seminal works on the few brightest, 
because lensed, quasars (Chartas et al. 2002, 2003, 2007), new observations of non-lensed QSOs with higher enough S/N 
are now starting to emerge. Two interesting examples are: i) HS 1700+6416 (z=2.73), 
the fourth high-z quasar that displays X-ray signatures of variable high-velocity outflows (UFOs), which is the only one that is not lensed (Lanzuisi et al. 2012); and ii) results 
of a multi-epoch observational campaign with XMM-Newton on the mini-BAL QSO PG 1126-041 (z=0.06) which has shown complex X-ray spectral variability, on  
time scales of both months and hours, best reproduced by means of variable and massive ionized absorbers along the line of sight (Giustini et al. 2011). 
In particular the highest S/N part of the observation has shown clear evidence in the highest energy band (E$>$5 keV) for the presence of a 
ultra-fast outflow with velocity of $\sim$ 16500 km s$^{-1}$ variable on very short time scales of a few ks (see Fig. 2).

\begin{figure}[]
{\includegraphics[width=6cm,height=8cm]{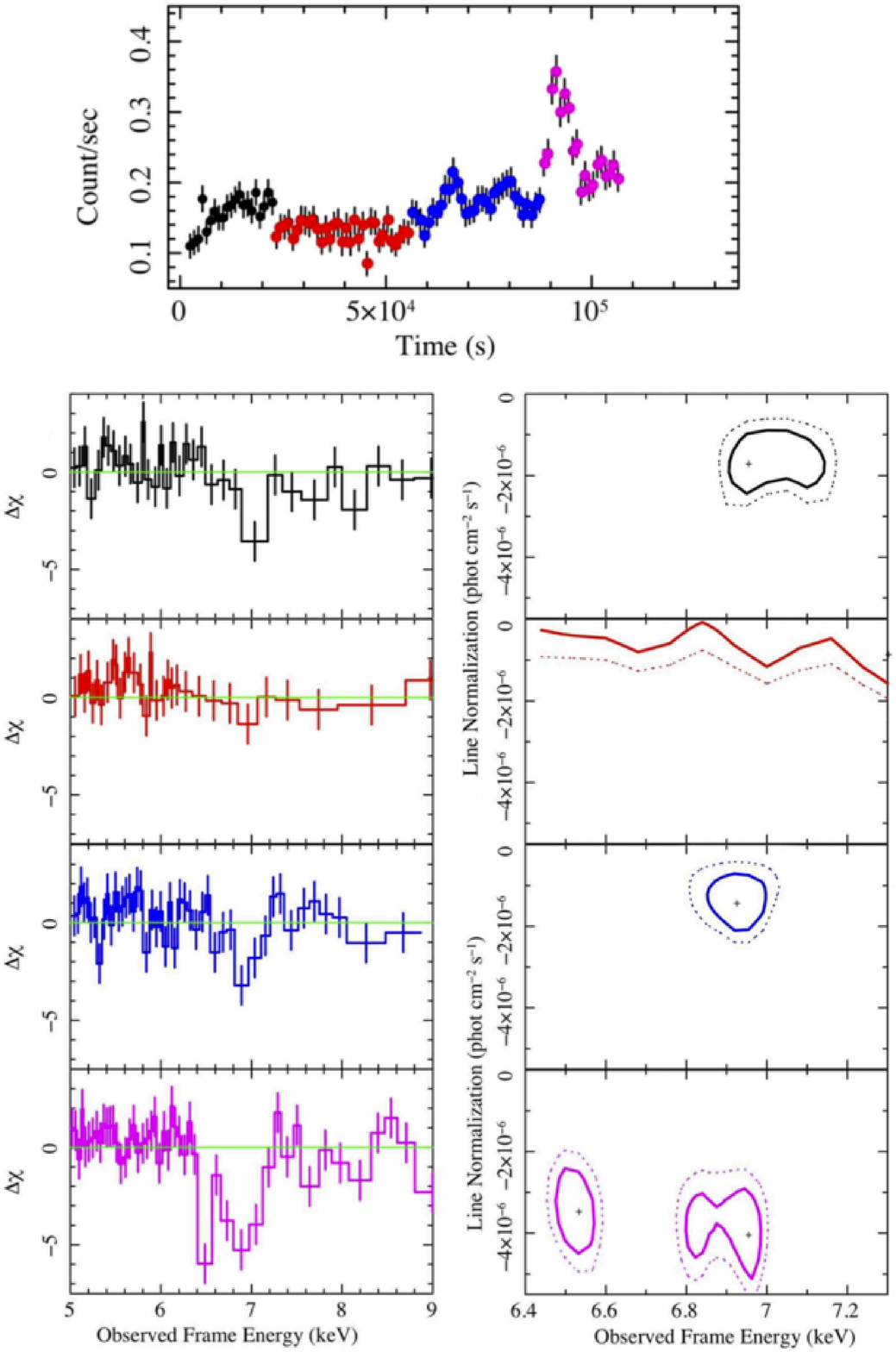}}
\caption{\footnotesize
(Top): 0.2-10 keV EPIC-pn light curve of PG1126-04. (Bottom): The left column shows the 5-9 keV EPIC-pn spectral residuals for the four (color-coded) time slices shown in the top panel. This evidences the presence of Fe K absorption variability over very short ($\sim$ks) time scales. The right column reports the 90\% and 68\% confidence levels for the energy and normalization of Gaussian absorption lines used to model the residuals in each slice. The strong variability of this highly-ionised (log $\xi$ $\sim$ 3.5 erg cm s$^{-1}$) and highly-massive (N$_W$$\sim$7$\times$ 10$^{23}$ cm$^{-2}$) phase of the wind over very short time scales (a few ks) is evident (from Giustini et al. 2011).
}
\label{eta}
\end{figure}

Even if the fraction of QSOs with UFOs remains to be determined yet, these findings are qualitatively consistent with AGN-driven accretion disk winds scenarios, and are pointing to a phenomenon which could be actually present/effective in a large fraction of all AGNs and QSOs. These observations have paved the way to, but also shown the need for, more sensitive observations and more time-resolved spectral analysis in the future.

\vspace{-0.3 cm}
\section{Understanding UFOs}

In an attempt to understand the global origin of these winds, in particular which of the possible physical mechanism(s) (heating, radiation, and/or magnetic fields)
is responsible for their formation and acceleration, Tombesi et al. (2013) have compared the properties of UFOs to those of lower ionization and 
lower velocities WAs. Their observed properties ($\xi$, N$_{W}$, and velocities), as well as the rough estimates for their 
distance, mass outflow rate and kinetic energy, span over several order of magnitudes. This allows the authors, for the very first time, to compare 
their global properties to theoretical calculations and tentatively derive a unifying picture.
Remarkably, all the absorbers are found to populate roughly continuously the whole parameter space, with UFOs observable closer to the central black hole, with higher ionization, 
column density, outflow velocity and mechanical power than WAs. These evidences strongly suggest that these absorbers, often considered of different types, 
could actually represent parts of a single large-scale stratified outflow, where UFOs are likely launched from the inner accretion disk while the WAs at larger distances, such as the outer 
disk and/or the putative obscuring torus (see Fig. 3). So far, the observed correlations among different parameters - and their slopes - appear to be consistent 
with radiation pressure through Compton scattering or MHD processes being responsible for the outflow acceleration. But there are also tempting indications that both processes 
could be acting simultaneously, with the MHD regime playing a substantial role (Tombesi et al. 2013). 
Clearly, more data and better measurements are needed for a more detailed and complete comparison with theoretical expectations (i.e. Sim et al. 2010; 
Fukumura et al., 2010, Kazanas et al. 2012).

\begin{figure}[t!]
{\includegraphics[width=7cm,height=2cm]{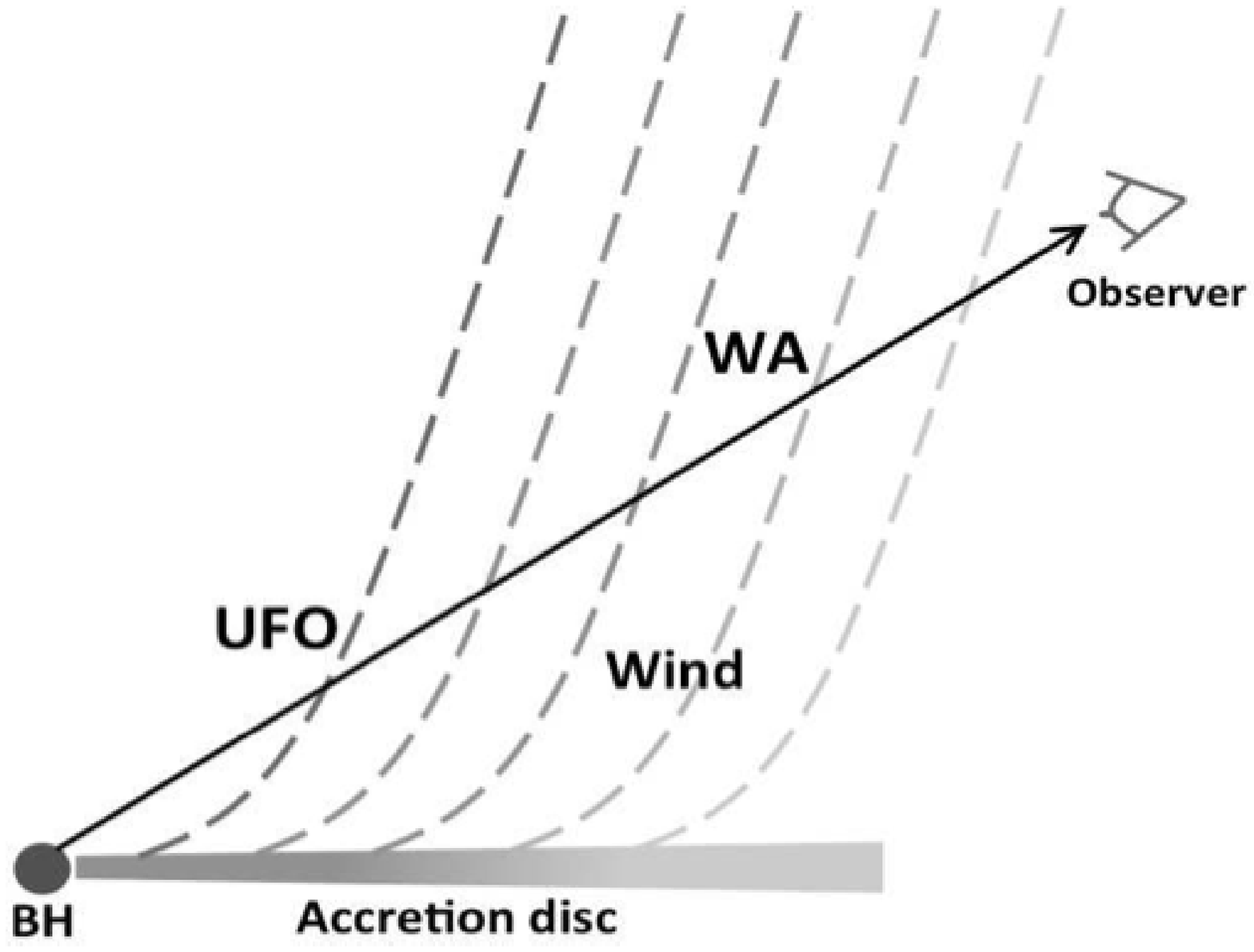}}
{\includegraphics[width=6cm,height=2.5cm]{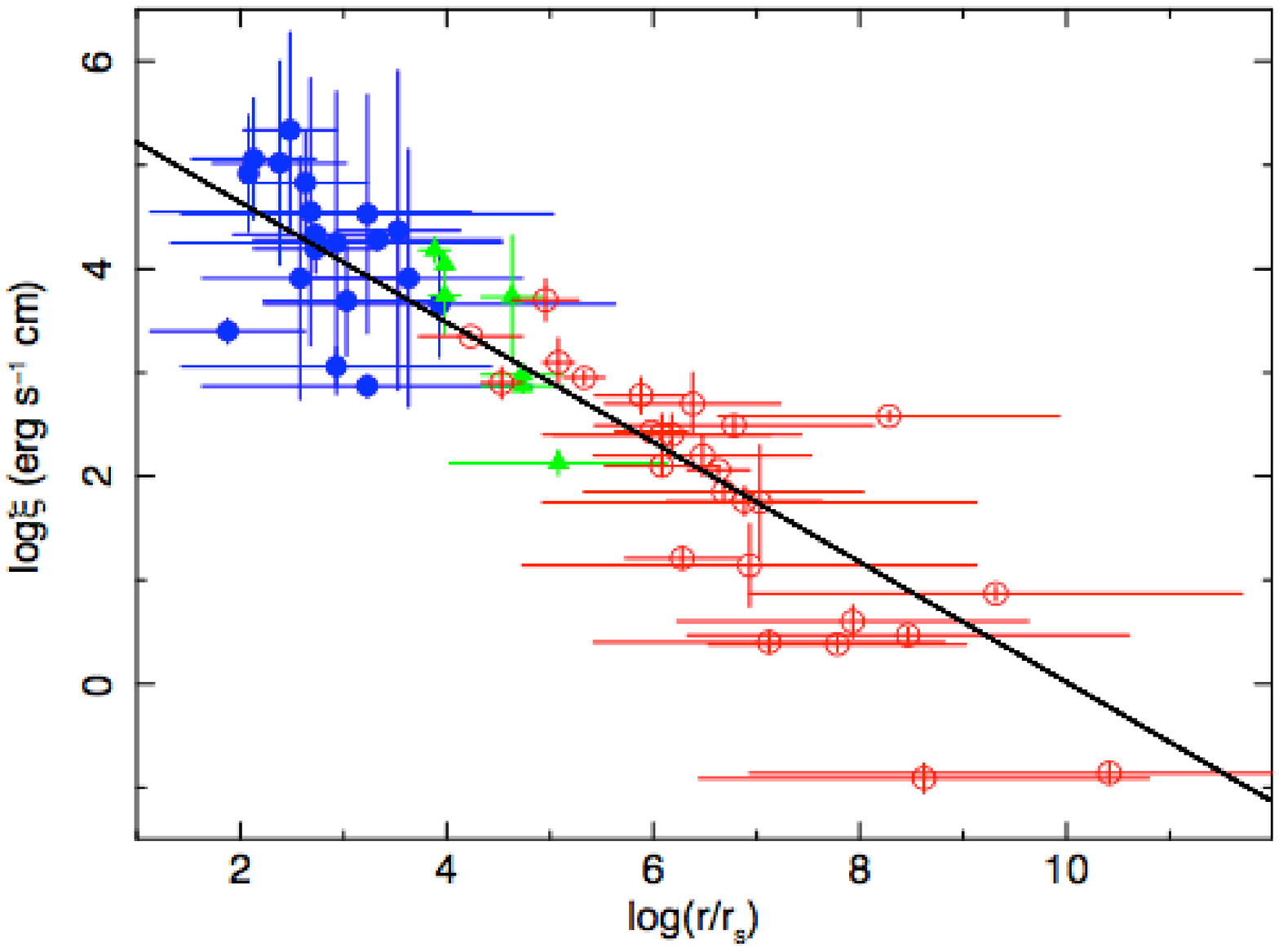}}
{\includegraphics[width=6cm,height=3cm]{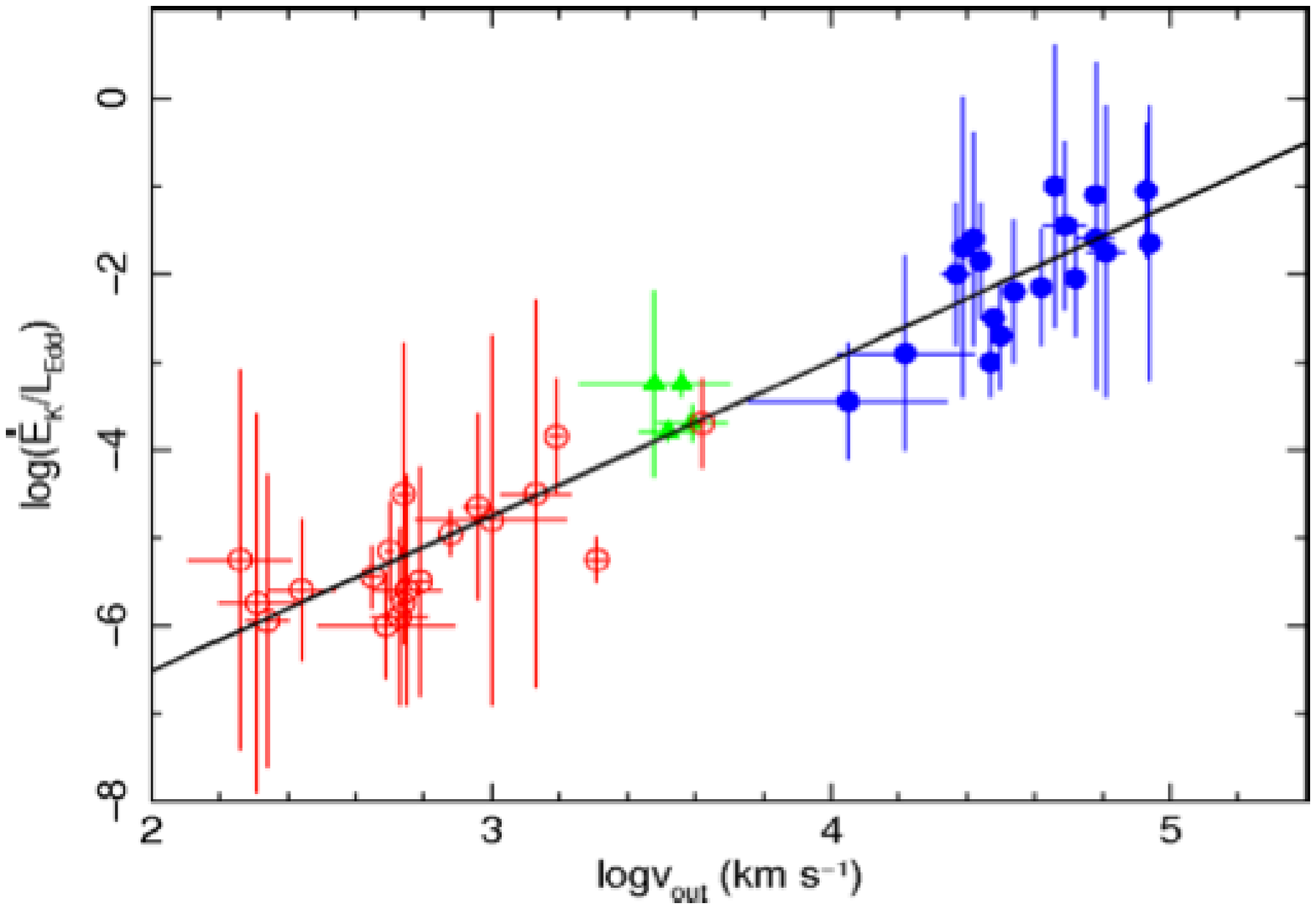}}
\caption{\footnotesize
(Top): Schematic view of an accretion disk wind by an observer located top right and looking through the wind, thereby probing UFOs/WAs absorption features from the inner/outer parts of the wind.
(Middle): Absorber ionization state vs. distance from the central source (in units of Schwarzschild radius). The blue, green and red points indicate the highly ionized UFOs, the intermediate ionization absorbers 
and the lower ionization WAs, respectively. The solid line represents the best-fit linear correlation curve. (Bottom): Kinetic power vs. outflow velocity (figures from Tombesi et al. 2013)}
\label{eta}
\end{figure}

\vspace{-0.3 cm}
\section{Impact on AGN Feedback}
As a "by-product" of the above studies, and expanding significantly upon other attempts (i.e. Crenshaw \& Kraemer, 2012), the potential global impact of 
winds (from WAs to UFOs) to AGN feedback was also estimated somewhat quantitatively in Tombesi et al. (2013). This work has shown that most of the outflows, especially the UFOs, 
are able to provide a mechanical power of at least $\sim$0.5\% and up to several\% of the AGN bolometric luminosity (Fig. 4), thereby releasing considerable feedback of energy and
momentum into the interstellar medium of the host galaxy. Stimulated by these findings, recent 3-D grid-based hydrodynamical simulations 
by Wagner et al. (2013) have demonstrated that, on kpc scales, feedback by UFOs operates similarly to feedback by relativistic AGN jets, and that considering the common occurrence of UFOs in AGN, they are likely to be important in the cosmological feedback of galaxy formation. Whether these outflows could also contribute significantly 
to energize and enrich the intergalactic medium of groups and/or clusters remains to be understood (but see Gaspari et al. 2012)

\vspace{-0.3 cm}
\section{Conclusions}

There have been significant advances in the last few years in characterizing the global properties of AGN winds, in particular from their most "extreme" components 
probed by their X-ray absorption features. Evidences of UFOs in nearby AGNs have become robust, and some are emerging in an increasing number of QSOs too. 
Some recent studies based on both observations and theoretical simulations have just started tackling the physical mechanisms at work in accelerating these outflows. 
Other studies have also demonstrated that UFOs could play a major role in providing enough momentum and energy feedback 
to quench the galaxy star formation. In the near-future, X-ray observations with micro-calorimeters (ASTRO-H) will certainly reduce further current uncertainties in the 
UFOs parameter-space. Multi-frequency observations and monitoring of nearby AGNs will also be clue in understanding the evolution of the nuclear (sub-pc) UFOs 
into the larger (pc and kpc) scales.
 
 \begin{figure}[]
{\includegraphics[width=6cm,height=3cm]{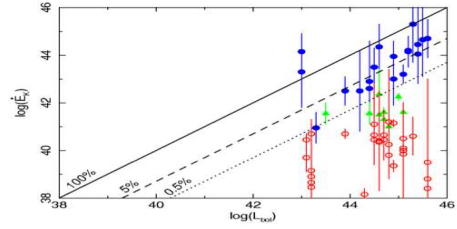}}
\caption{\footnotesize
Outflows kinetic power with respect to the AGN bolometric luminosity (from Tombesi et al. 2013). The points correspond to the WAs (red open circles), non-UFOs (green filled triangles) and UFOs (blue filled circles), respectively. The transverse lines indicate the ratios between the outflow mechanical power and bolometric luminosity of 100\% (solid), 5\% (dashed) and 0.5\% (dotted), respectively.
}
\label{eta}
\end{figure}


\vspace{-0.5 cm}
\bibliographystyle{aa}

\end{document}